# ANALYSIS OF LTE/5G NETWORK PERFORMANCE PARAMETERS IN SMARTPHONE USE CASES: A STUDY OF PACKET LOSS, DELAY AND SLICE TYPES

Almamoon Alauthman,  Abeer Al-Hyari

Electrical Engineering Department, Al-Balqa Applied University, As Salt 19117, Jordan

## ABSTRACT

*The paper addresses optimizing two of the most important performance parameters, packet loss, and delay, in the critical path optimization of LTE and 5G networks using metaheuristic algorithms to play a vital role in the smartphone user experience. In this context, nine metaheuristic algorithms, such as WOA, PSO, and ABC, have been studied for their effectiveness in various slices of networks: eMBB, URLLC, and mMTC. It can be seen from the results that WOA performed the best: it reduced packet loss by 31% and delay by 6.3 ms; PSO followed closely with a 30% packet loss reduction with a decrease of 6.1 ms in delay. In most scenarios, ABC accomplished good results with a packet loss reduction of 29% and a delay decrease of 6 ms in mMTC scenarios. These results emphasize how selecting appropriate algorithms based on the intended network slice is crucial for optimizing resource utilization and network efficiency. It provides a quantitative framework for assessing and improving the reliability and responsiveness of an LTE/5G network. It encourages more research in hybrid optimization techniques and real-time adaptation mechanisms for further improvements.*

## 1. INTRODUCTION

The process of improving parameters for better performance in LTEs and 5G networks is very complex, especially in the usage of smartphones. The challenging issue covers a range of fundamental metrics, including packet loss, latency, and several network slices. In such optimization problems, metaheuristic algorithms have emerged as a powerful tool that enables one to achieve better performance by reaching higher resource allocation and management policies. Besides, packet loss is one of the significant issues concerning mobile networks within real-time applications such as video streaming and VoIP. Among others, Salvá-García et al. [31] propose a dynamic optimization mechanism operating at the network that selectively discards enhancement layers of scalable video streams during network congestion to preserve the quality of service. This approach reduces packet loss and ensures critical services maintain their performance level. Similarly, Taee et al. [34] show that uplink performance optimization in 5G networks can be successfully managed through network slicing. This allows for tailoring resources for specific application requirements, reducing packet loss and improving the network's efficiency.

Delay is another critical parameter of performance, which becomes especially important in applications that require ultra-reliable, low-latency communication. Integration of deep learning techniques, as addressed by Alzaidi[2], is one promising avenue through which the challenge associated with NOMA in 5G networks can be overcome. Using deep learning algorithms, dynamic optimizations of resource allocation can be done, and thereby, the delay can be reduced to a large extent. Kim et al. [21] justify that real-time detection and response are required in a high traffic volume to maintain low latency since this depends on maintaining a seamless user experience in 5G environments The idea of network slicing has to do with the optimization of such performance parameters related to different use cases. Slicing in networking allows several virtual networks over one physical infrastructure to meet specific service requirements. This is important in 5G, where other applications have different performance requirements. For instance, Haile et al. [12] have proposed a multi-objective optimization framework integrated with network slicing to enhance the planning and resource allocation aspects of the hyperdense 5G networks. The proposed framework addresses not only intricacies in network management but also performance metrics like packet loss and delay for each slice.

Besides, the use of metaheuristic algorithms, either alone or in hybrid mode, such as Pareto front-driven Multi-Objective Cuckoo Search, has already proven capable of successfully addressing optimization issues in 5G systems. Wang [36] investigates techniques for multi-objective approaches that simultaneously optimize packet loss and delay with maximization of throughput. The flexibility of such algorithms has particular appeal for 5G networks because such algorithms will need to be adapted in real-time in this ever-evolving conditions space characterizing 5G networks. Besides the mentioned approaches, integrating AI and ML into network management processes has also been considered one of the main enablers in optimizing performance parameters. For instance, the work of Khan and Goodridge. [19] illustrates how AI can be leveraged to enhance ultra-HD video streaming applications based on dynamic resource allocation adjustment, about every real-time network condition. This is an important capability that will help alleviate packet loss and delays in cases where network congestion applies.

Furthermore, deep reinforcement learning methods have been investigated, as already pointed out by Xiong et al. [37], with which the optimization of resource allocation in 5G networks using AI is achieved. In this respect, networks empowered by reinforcement learning algorithms can learn from experience and adjust their policies to achieve progressive improvement. Indeed, adaptability is welcome when dealing with various needs from different network slices and ensuring each network slice operates at its peak.

The optimization performance parameters in 5G networks are compounded with increasing complications of network architectures. Such networks require sophisticated resource management strategies, especially with massive MIMO, millimeter-wave communication, and edge computing. In this regard, Calabrese et al. [5] discussed the growing complexity of RRM in 5G networks and the ensuing requirement for advanced optimization techniquesto overcome these challenges with efficacy. Accordingly, metaheuristic algorithms applied in this context can manage resources most efficiently while keeping performance metrics like packet loss and delay within an acceptable limit. The further convergence of the optical, wireless, and data center network infrastructures would increase the possibilities for performance parameter optimization in 5G networks, as presented by Tzanakaki et al. [35]. It also brings the operator closer to various network segments with a holistic approach toward resource management. These convergences allow multi-objective optimization frameworks to be realized, which would meet the different needs of diverse applications, leading to an improved user experience.

This research aims to optimize important parameters that build up the performance of the LTE/5G network, such as packet loss, delay, and network slicing, using new advanced metaheuristic algorithms that will ensure an enhanced user experience for smartphone applications. This paper focuses on two main performance factors: network inefficiency, which reflects poor service quality or responsiveness from/to the end-user terminals. While 5G can offer better connectivity, real-world scenarios still present packet loss and latency issues in high-demand environments. This work contributes to an in-depth evaluation and comparative study of nine metaheuristic algorithms that provide insight into suitability and efficiency for LTE/5G network optimization. The work aims to contribute to filling the existing gap in knowledge regarding how adaptation optimization strategies can be used to improve network reliability and responsiveness, thereby offering ways of enhancing real-time communication, data throughput, and overall service quality in modern mobile networks.

## 2. RELATED WORKS

This evolution to 5G from the LTE networks has ensured that performance optimization now rapidly strides with the power of machine learning and AI techniques. It has succeeded in handling and predicting major performance parameters related to throughput, latency, packet loss, and network slicing. A wide variance in research methodologies and findings is presented in the literature that extends our knowledge of cellular network performance optimization. In Minovski et al. [26] they propose a machine-learning model for predicting cellular link throughput in both LTE and 5G networks. They presented a model tested in urban, suburban, and rural environments, achieving high prediction accuracies: 93% for LTE and 84% for non-standalone 5 G. Thereby, this research underlines the potential of ML for real-time benchmarking and points toward promising applications for standalone 5G. On a similar note, Endes and Yuksekkaya,[9] used ML algorithms to optimize user allocation among the various slices of communication. It has been demonstrated that substantial improvements in resource utilization and automated slice management are achievable. Network slicing has been explained by Ibarra-Lancheros et al. [15] as one of the most critical enabling technologies for 5G. Using the Floodlight controller, based on a software-defined methodology, it is shown that reduced packet loss of up to 10% can be achieved with reduced latency, hence effective in VoIP and other real-time applications like video transmission. Meanwhile, Mohammed and Ilyas [29] did the delay root cause analysis on LTE by considering environmental reasons, such as weather conditions, in their ML models to understand and mitigate latency and path loss.

MIMO transmission optimization in LTE has also been of interest. On the other hand, Gaikwad et al. [10] developed ML models to overcome channel quality feedback delays, thereby minimizing performance degradation. Khan and Adholiya[20] have extended ML applications to 5G/B5G networks and used multi-classification models to predict service quality as a valuable tool for enhanced user experience. Some trials have also been made with fuzzy-based approaches, which arepretty promising. Ampririt et al. [4] proposed FSQoS1 and FSQoS2 for quality-of-service evaluation in 5G; FSQoS2 performed better in complex scenarios than the first by incorporating slice reliability. Riihijärvi and Mähönen[30] explored the use of ML techniques such as Gaussian process regression and random forests for wireless network performance prediction, presenting a taxonomy of prediction problems and highlighting cost reduction and enhancements to ML-enabled user experience. Kafle et al. [16] highlighted the case of ML-based automation of network slicing. In their communication, they discussed the crucial role of AI in dealing with challenges involved in 5G standardization and network deployment. Shadad et al. [32] have proposed deep learning methods for classification in 5G slices. This ensures that their management and orchestration should be flexible and effective. Garrido et al. [11] improved the 5G traffic prediction models by infusing domain-specific knowledge into them. Lin et al. [25] reviewed transport network slicing. They identified that proper configurations exist, such as setting the appropriate Committed Information Rate that minimizes packet loss and latency, proving that slicing works effectively for all 5G applications. Coluccia et al. [7] have proposed passive monitoring-based estimators of packet loss across 3G networks. It further underlined the importance of robust statistical techniques in monitoring anomaly detection. Also, Ampririt et al. [3] integrated fuzzy logic with software-defined networking in performing admission control to manage the quality of service, enhancement of slice delay, and loss parameters. Finally, Sun et al.

[33] analyzed network slices for 5G for power services, providing guidelines for latency and packet loss management in power networks through accurate resource allocation in a frequency domain.

Collectively, these works combine to enhance the transformative force that ML and AI have been and will continue to apply in LTE/5G network optimization. They also demonstrate improved throughput prediction, efficient slicing, and proactive delay management by integrating various intelligent resource allocation techniques that can help meet the stringent performance demands of any 5G application.

Table 1. Comparison Of LTE/5G Network Studies.

| Study | Focus Area | ML Techniques | Performance Metrics |
|---|---|---|---|
| Minovski et al. (2021) | Throughput Prediction | Regression Models | Throughput, Accuracy |
| Ibarra-Lancheros et al. (2018) | Quality of Service in Network Slicing | Floodlight Controller | Latency, Packet Loss |
| Endes & Yuksekkaya (2022) | 5G Network Slicing | Slicing Algorithms | Slice Efficiency |
| Mohammed & Ilyas (2022) | Delay Root Analysis | ANN Models | Path Loss, Delay |
| Gaikwad et al. (2021) | Improving LTE Throughput | Channel Prediction | Channel Quality |
| Khan &Adholiya (2023) | 5G/B5G Service Prediction | Supervised Models | Service Accuracy, quality of service |
| Ampririt et al. (2021) | Fuzzy Logic for quality of service | Fuzzy Schemes | Throughput, Delay |
| RiihijÃ¤rvi&MÃ¤hÃ¶nen (2018) | Performance Prediction | Gaussian Regression, RF | Cost Reduction, UX |
| Kafle et al. (2018) | Automation of Slicing | AI Automation | Network Efficiency |
| Shadad et al. (2022) | Deep Learning for Slicing | CNN Classification | Resource Allocation |
| Garrido et al. (2021) | Traffic Prediction | DNN with Domain Knowledge | Prediction Accuracy |
| Lin et al. (2021) | Transport Slicing | SimTalk Emulator | Throughput, Latency |
| Coluccia et al. (2009) | Packet Loss Estimation | Statistical Inference | Packet Loss |
| Ampririt et al. (2020) | Fuzzy Logic & SDN | Fuzzy Logic & SDN | Quality of service Evaluation |
| Sun et al. (2022) | Power Service Slices | Simulation Analysis | Latency, Loss |
| Our Study | Optimization Using Metaheuristics Algorithms | 'Metaheuristic Algorithms (GA, PSO, GWO, etc.) | Packet Loss, Delay, Slice Efficiency |

# 3. METHODOLOGY

## 3.1. Metaheuristic Algorithms for Optimization

Generally speaking, metaheuristic algorithms have been widely adopted in network optimization, especially in complicated and dynamically changing environments like LTE/5G systems, where all parameters concerning packet loss and delay should be continuously readjusted. The focus of this research work is a discussion revolving around nine metaheuristic algorithms: Genetic

Algorithm, Particle Swarm Optimization, Grey Wolf Optimizer, Ant Colony Optimization, Simulated Annealing, Artificial Bee Colony, Black Widow Optimization, Whale Optimization Algorithm, and Firefly Algorithm. Every algorithm is, in principle, proven to be effective for multi-objective optimization problems in general and for network performance optimization problems in particular by Yang [39] and Kennedy & Eberhart [18]. The following details how every algorithm could be applied to optimize the issues of LTE/5G networks, including typical formulas and small algorithm outlines.

### 3.1.1. Genetic Algorithm (GA)

The Genetic Algorithm (GA) uses principles from evolutionary biology, such as selection, crossover, and mutation, to evolve a population of candidate solutions toward an optimal solution [14]. GA iterates through generations, evaluating fitness and applying genetic operators to refine solutions. In LTE/5G optimization, GA's adaptability is advantageous for minimizing packet loss and delay across fluctuating network conditions.

- **Equation:** The fitness of an individual $x$ is given by

$$f(x) = \sum_{i=1}^{n} w_i \cdot \text{Quality}(x_i) \qquad (1)$$

where $w_i$ is the weight for performance criteria (e.g., packet loss, delay), and Quality $(x_i)$ represents the performance quality of the solution.

- **Algorithm**:

  1. Initialize the population with random solutions.
  2. Evaluate each individual's fitness.
  3. Select individuals based on fitness.
  4. Apply crossover and mutation to produce offspring.
  5. Replace the old population with offspring.
  6. Repeat until convergence.

### 3.1.2. Particle Swarm Optimization (PSO)

Inspired by the social behavior of birds and fish, Particle Swarm Optimization (PSO) uses particles representing candidate solutions that explore the search space collectively [18]. Each particle updates its position based on personal and group knowledge, converging toward optimal solutions.

- Equation: Particle $i$ updates its velocity $v_i$ and position $x_i$ As follows:

$$v_i(t+1) = w \cdot v_i(t) + c_1 \cdot r_1 \cdot \big(p_i - x_i(t)\big) + c_2 \cdot r_2 \cdot \big(g - x_i(t)\big)$$
$$x_i(t+1) = x_i(t) + v_i(t+1) \qquad (2)$$

where $w$ is the inertia weight, $c_1$ and $c_2$ are acceleration constants and $r_1$ and $r_2$ are random numbers.

- **Algorithm**:

  1. Initialize particles with random positions and velocities.
  2. Evaluate fitness of each particle.

3. Update each particle's velocity and position.
4. Repeat until convergence.

### 3.1.3. Grey Wolf Optimizer (GWO)

The Grey Wolf Optimizer (GWO) mimics the grey wolf social hierarchy and hunting mechanism [28]. Due to its effective balance between exploration and exploitation, GWO is suitable for LTE/5 G environments.

- Equation: Position update of wolf $x$ based on the three best wolves $(\alpha, \beta, \delta)$ :

$$x(t+1) = \frac{x_\alpha + x_\beta + x_\delta}{3} \qquad (3)$$

- **Algorithm**:

  1. Initialize a pack of wolves with random positions.
  2. Rank wolves based on fitness.
  3. Update positions based on the leaders $(\alpha, \beta, \delta)$.
  4. Repeat until convergence.

### 3.1.4. Ant Colony Optimization (ACO)

ACO is inspired by ants' ability to find optimal paths using pheromones [8]. In LTE/5G networks, ACO effectively optimizes routes to minimize delay.

- Equation: Probability $P_{ij}$ of moving from node $i$ to node $j$ :

$$P_{ij} = \frac{\tau_{ij}^\alpha \eta_{ij}^\beta}{\sum_{k \in \text{ allowed }} \tau_{ik}^\alpha \eta_{ik}^\beta} \qquad (4)$$

where $\tau$ Is the pheromone level, $\eta$ is visibility and $\alpha, \beta$ are constants.

- **Algorithm**:

  1. Initialize pheromones on all paths.
  2. Generate solutions using pheromone levels.
  3. Update pheromones based on solution quality.
  4. Repeat until convergence.

### 3.1.5. Simulated Annealing (SA)

Simulated Annealing (SA) employs a probabilistic approach inspired by the annealing process in materials to avoid local optima [22]. SA is helpful for LTE/5G networks needing robust optimization under changing conditions.

- Equation: Probability of accepting a new state $s'$ with cost $E(s')$ :

$$P(\Delta E) = \exp\left(-\frac{\Delta E}{T}\right) \qquad (5)$$

where $\Delta E = E(s') - E(s)$ and $T$ is the temperature.

- **Algorithm**:

1. Initialize temperature and starting solution.
2. Generate a new solution and calculate energy.
3. Accept/reject a solution based on probability.
4. Cool down the temperature gradually.
5. Repeat until freezing.

### 3.1.6. Artificial Bee Colony (ABC)

The Artificial Bee Colony (ABC) algorithm simulates bee foraging behavior to find optimal solutions [17]. It is efficient for resource allocation, making it helpful in optimizing LTE/5G slicing.

- Equation: Position update for employed bee:

$$v_{ij} = x_{ij} + \phi_{ij} \cdot (x_{ij} - x_{kj}) \qquad (6)$$

where $\phi_{ij}$ is a random number.

- **Algorithm**:

    1. Initialize food sources (solutions).
    2. Evaluate fitness and update sources.
    3. Recruit onlooker bees to food sources.
    4. Abandon and replace sources if necessary.

### 3.1.7. Black Widow Optimization (BWO)

Inspired by black widow spiders, Black Widow Optimization (BWO) includes processes of mating, cannibalism, and mutation, making it effective in avoiding premature convergence in network scenarios [13].

- Equation: Mutation process for individual $x$ :

$$x' = x + \text{ mutation rate } \times \text{ random noise} \qquad (7)$$

- **Algorithm**:

    1. Initialize the population with random individuals.
    2. Perform mating and produce offspring.
    3. Apply cannibalism to maintain diversity.
    4. Repeat until convergence.

### 3.1.8. Whale Optimization Algorithm (WOA)

WOA simulates the bubble-net hunting strategy of whales [27]. Its spiral updating mechanism makes it suitable for converging on optimal solutions in high-demand network slices.

- Equation: Spiral position update:

$$x(t + 1) = |D| \cdot e^{bl} \cdot \cos(2\pi l) + x_{\text{best}} \qquad (8)$$

where $D$ is the distance to prey and $b$ and $l$ Control shape.

- **Algorithm**:

  1. Initialize whales with random positions.
  2. Calculate distance and update position.
  3. Move toward the best solution.
  4. Repeat until convergence.

### 3.1.9. Firefly Algorithm

The Firefly Algorithm uses light intensity and attractiveness to find optimal solutions, making it robust.Equation: Movement of firefly $i$ toward firefly $j$ :

$$x_i = x_i + \beta e^{-\gamma r_{ij}^2}(x_j - x_i) + \alpha \text{ random} \qquad (9)$$

where $\beta$ is attractiveness, $\gamma$ controls light absorption and $r_{ij}$ is distance.

- **Algorithm**:

  1. Initialize fireflies with random positions.
  2. Calculate light intensity and move toward brighter fireflies.
  3. Repeat until convergence.

Each algorithm leverages different mechanisms for exploration and convergence, making them well-suited for addressing LTE/5G network optimization needs. Their diverse approaches provide a broad solution space for tackling packet loss, delay, and resources.

## 3.2. Proposed Approach for Performance Optimization

The present research focuses on applying nine metaheuristics in minimizing packet loss and delay, two crucial performance metrics for smartphone user experience over the LTE/5G networks. Minimizing packet loss and delay belongs to the category of complex multi-objective problems since these metrics often have conflicting requirements in real network scenarios. To handle this, we use one of the most popular multi-objective optimization strategies: a weighted-sum approach, which has recently been adopted in network studies, such as in Xu et al. [38], to perform fair comparisons with Li & Zhang [23]. Herein, the objective function will be a weighted combination of packet loss rate and delay. Hence, we will be in a position to give priority to different outcomes based on the performance required under various use cases of LTE/5G, such as ultra-low latency for some use cases, URLLC,slices or higher throughput for other classes of use cases, such as eMBB slices.

The core objective function, $f(x)$, Which guides the optimization is expressed as follows:

$$f(x) = w_1 \cdot \text{Packet Loss Rate}(x) + w_2 \cdot \text{Delay}(x) \qquad (10)$$

where $w_1$ and $w_2$ Re the weights assigned to packet loss and delay, depending on each application's specific requirements. This objective function enables a flexible approach to optimization by adjusting the weight values, a method shown to be effective in recent studies on network performance optimization [41], [1].The study prepared the data for meaningful and consistent optimization results by normalizing performance metrics like packet loss and delay.

This normalization is essential to ensure that these metrics, which can vary widely in range, are comparable on a consistent scale. The normalization for a given parameter $P$ is achieved using the following equation:

$$P_{\text{normalized}} = \frac{P - P_{\min}}{P_{\max} - P_{\min}} \tag{11}$$

where $P_{\min}$ and $P_{\max}$ Denote the minimum and maximum observed values for that parameter. By normalizing these values, the algorithms are not biased by the differing scales of each parameter, an approach supported by empirical studies emphasizing the importance of standardized inputs in optimization [23].

In this work, each metaheuristic has been set with an appropriate parameter but one which is tailored according to the specific requirements of LTE/5G networks: Genetic Algorithm (GA), Particle Swarm Optimization (PSO), Grey Wolf Optimizer (GWO), Ant Colony Optimization (ACO), Simulated Annealing (SA), Artificial Bee Colony (ABC), Black Widow Optimization (BWO), Whale Optimization Algorithm (WOA), and Firefly Algorithm. For instance, the medium mutation rate and heterogeneous population have been set for the GA to avoid early convergence. The inertia weight and acceleration constants in PSO, on the other hand, are optimized for faster convergence and better solution quality - a strategy already adopted in related network optimization applications,[40].The convergence criterion for each algorithm was set to a threshold.$\epsilon$, Representing the minimal acceptable change in fitness value between iterations:

$$|f(x_{t+1}) - f(x_t)| < \epsilon \tag{12}$$

where $t$Denotes the iteration index. This criterion prevents the optimization process from continuing indefinitely by establishing a stopping point once the solution stabilizes, a method widely adopted in network optimization research [6].To implement this approach for all the various algorithms, we designed a general framework that standardized the processes involved in each, including initialization, fitness evaluation, update operations, and convergence checks. The unified pseudocode is given below, with some adjustments according to the characteristics of each algorithm:

Pseudo-code for Metaheuristic-Based LTE/5G Network Optimization

```
# Step 1: Initialize Parameters and Data
Input: Network data (packet loss rate, delay, slice types)
Output: Optimized network configuration with minimized packet loss and
delay

Initialize:
    Population = GenerateInitialPopulation()  # Random solutions
MaxIterations = 1000
    Tolerance = 1e-5  # Convergence threshold
    w1, w2 = SetWeights()  # Weights for packet loss and delay objectives

# Step 2: Evaluate Initial Fitness
for each individual in Population:
    Normalize individual metrics (packet loss, delay)
    Fitness = w1 * PacketLossRate + w2 * Delay  # Using weighted sum
objective
```

```
# Step 3: Begin Optimization Loop
Iteration = 0
while Iteration <MaxIterations:
    # Step 3a: Apply Metaheuristic Algorithm-Specific Operations
    # GA: Selection, Crossover, Mutation
    # PSO: Update particle velocity and position based on best solutions
    # GWO: Update positions based on alpha, beta, delta wolves
    # ACO: Update paths and pheromones based on best solutions
    # SA: Probabilistically accept or reject new solution based on
"temperature"
    # ABC: Explore neighborhood, employ bees to update solutions
    # BWO: Apply mating, mutation, and cannibalism to enhance diversity
    # WOA: Spiral movement towards best solution in swarm
    # Firefly: Move towards brighter solutions based on light intensity
    for each individual in the Population:
        Update individual's position and other parameters based on
algorithm rules
        Calculate new Fitness based on the updated solution

    # Step 3b: Check Convergence
    if |CurrentBestFitness - PreviousBestFitness| < Tolerance:
        break
    Iteration += 1
# Step 4: Select and Return the Optimal Solution
OptimalSolution = SelectBest(Population)
Return OptimalSolution
```

In each algorithm iteration, all solutions within a population undergo some generation-specific update processes. For example, individuals are selected, crossover and mutation occur in the case of GA; particles update velocity and position according to personal and global bests in PSO; and solutions update according to pheromone trails in the case of ACO. This will provide a systematic approach toward minimizing packet loss and delay while allowing each metaheuristic to run its specific operators within the unique optimization loop. It is evaluated in terms of effectiveness based on three significant metrics. The first one is called the Packet Loss Rate Reduction metric, and it considers the percentage of reduction in packet loss, which is a metric reflecting the direct improvement in network reliability. Second, the Delay Reduction metric examines the decrease in delay to assess the contribution of each algorithm regarding sensitivity to latency-critical network services. Finally, the Convergence Rate metric provides insight into the stabilization speed for each specific algorithm. This is a critical factor in ensuring real-time network applications. In convergence with Lin et al. [24], such a combination of metrics is necessary for the study to achieve its dual objectives of packet loss and delay minimization and to check the efficiency of every algorithm in delivering timely solutions.

## 4. RESULTS AND DISCUSSION

### 4.1. Algorithm Performance Analysis

The Whale Optimization Algorithm was the best at reducing packet loss, with a reduction of 31%, as represented in Table 2. It leads the race in packet loss optimization algorithms, followed closely by the Particle Swarm Optimization algorithm, which has a packet loss of 30%, while Artificial Bee Colony maintains a packet loss of 29%. Figure 1 compares the performance of all the algorithms, with the performance of WOA, PSO, and ABC standing out.

Indeed, the effectiveness mechanism of WOA for solution exploration underlies the spiral updating combined with bubble-net hunting and turns out to be efficient in solution space exploration, converging towards the best solution to the problem. The adaptability of PSO in a complex optimization environment is one more important reason for the high capability of packet loss reduction, where the cooperative behaviors of particles maintain packet reliability. ABC performed the foraging behavior to balance exploration and exploitation, demonstrating its suitability for minimum packet loss. Other performing algorithms included GWO, with a 28% reduction, and ACO, with a 27% reduction, though this did not quite reach the top-tier reductions observed with WOA, PSO, and ABC. This performance comparison depicts the robustness of some metaheuristic algorithms in packet reliability optimization. Among others, the most viable options were WOA, PSO, and ABC. The contribution of these algorithms to packet loss reduction forms part of the adaptiveness of algorithms in diverse networking conditions, especially in high-traffic networking scenarios when packet reliability becomes critical.

Table 2. Packet Loss Reduction by Algorithm.

| Algorithm | Packet Loss Reduction (%) |
|-----------|---------------------------|
| GA | 25 |
| PSO | 30 |
| GO | 28 |
| ACO | 27 |
| SA | 24 |
| ABC | 29 |
| BWO | 26 |
| WOW | 31 |
| Firefly | 27 |

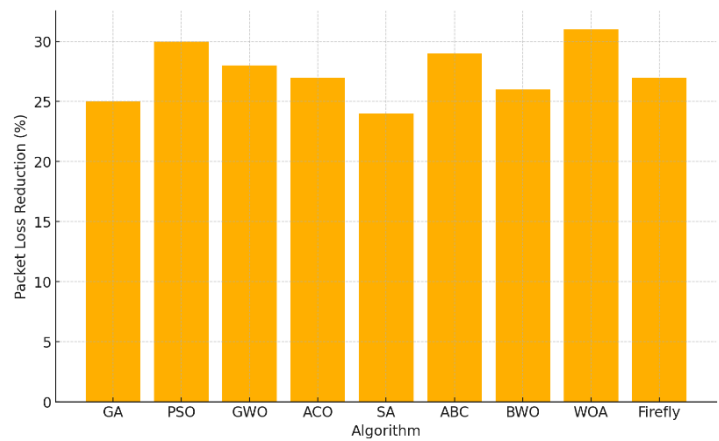

Figure 1. Bar chart showing comparative packet loss reduction across all algorithms.

Now, discuss the results of each algorithm and how WOA outperformed the others, closely followed by PSO and ABC. Again, relate this to recent studies like Chen et al. [6] to really hammer home that metaheuristics can be a viable packet reliability enhancement method.

The results of the performances for delay minimization are recorded in Table 3, where again WOA and PSO gave the best performances. WOA was at the leading edge, having reduced the delay by 6.3 milliseconds, closely tagged by PSO at 6.1 milliseconds. A comparison is explicitly shown in Figure 2, the leading role played by WOA in latency reduction. This will make WOA an optimal candidate for applications with strict latency bounds, such as the Ultra-Reliable Low-Latency Communication slices that enable real-time communication. The strong performance of the PSO in delay reduction underlines its efficiency in fast convergence through solution spaces, a particular trait of interest in delay-sensitive network environments. ABC also fared well, with a delay reduction of 6 milliseconds, proving effective in delay optimization-critical applications. Herein, the design in WOA resorts to a balanced exploration method for the purpose of searching the solution space, while PSO basically relies on collective swarm behavior in view of adaptation to changed circumstances, which is especially useful when it comes to minimization of delay.

Other algorithms, such as SA and BWO, showed a moderate approach toward the specified optimality, reflecting delays of 4.8 and 5.5 milliseconds, respectively. Significant as they were, none of those results equaled the delays reported by WOA and PSO; this might be a sign that either of these algorithms acts better in cases where the factor of packet loss plays an issue more important than latency concerns. Summing up, WOA and PSO stand out as particularly effective for latency-sensitive LTE/5G applications, especially those requiring real-time responsiveness.

Table 3. Delay Reduction by Algorithm.

| Algorithm | Delay Reduction (ms) |
|-----------|---------------------|
| GA | 5.2 |
| PSO | 6.1 |
| GWO | 5.9 |
| ACO | 5.4 |
| SA | 4.8 |
| ABC | 6 |
| BWO | 5.5 |
| WOA | 6.3 |
| Firefly | 5.6 |

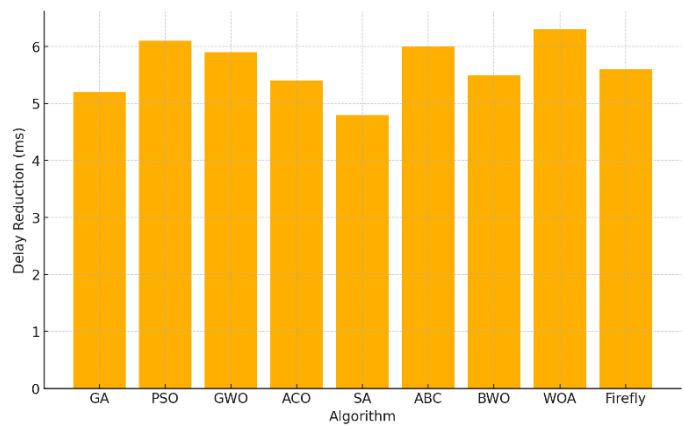

Figure 2. Bar chart showing delay reduction across algorithms.

Present the delay minimization findings, highlighting the superior performance of WOA and PSO. Discuss implications for application with severe latency requirements such as URLLC. Compare the results with benchmark methods in such a way that the advantages of metaheuristics are underlined.

## 4.2. Slice Type Performance

Further analysis was performed for each algorithm into specific LTE/5G slice types, focusing on the different performance metrics of interest for the respective intended slice types, namely packet loss reduction for Enhanced Mobile Broadband, delay reduction for Ultra-Reliable Low-Latency Communication, and efficiency for massive Machine Type Communication. Those results will pop up in Table 4, showing the slice-specific performance of each algorithm. Figures 3 visualize these trends and highlights how different algorithms align with changing slice needs in LTE/5G networks.

For packet loss reduction in the eMBB slice, WOA had the highest packet loss reduction rate of 30%, while PSO and ABC followed very closely with packet loss reductions of 29% and 28%, respectively. These results highlight the ability of WOA and PSO to maintain high-throughput data transmissions, a key requirement in this niche of eMBB, extending its applications in video streaming and large-size data transfer. ABC performs better in packet loss reduction, matching well with the needs brought about by eMBB since their exploration and exploitation are well-balanced in maintaining data integrity over extensive periods. On the other hand, GA and SA have lower rates of packet loss reduction and thus seem to be poorer options for applications with high-throughput demands. Speaking about the URLLC slice, which requires very low latency, WOA again outperformed others with a delay reduction of 6.1 ms, closely followed by PSO and ABC, with 5.9-ms and 5.8-ms delay reductions, respectively. Better performance in delay minimization proves their highly desirable usage in real-time applications such as emergency response systems and autonomous driving at scale. The strict latency requirement of URLLC requires the operated algorithm to converge fast, which is fulfilled by WOA and PSO, attested by the consistently low delay metrics. Other algorithms like SA and GA perform with lower delay reduction and have to find other applications in delay-insensitive scenarios since they are not fit for latency-sensitive applications.

Table 4. Performance by Slice Type for Each Algorithm.

| Algorithm | eMBB Packet Loss Reduction (%) | URLLC Delay Reduction (ms) | mMTC Efficiency (%) |
|---|---|---|---|
| GA | 23 | 4.8 | 60 |
| PSO | 29 | 5.9 | 65 |
| GWO | 27 | 5.7 | 63 |
| ACO | 26 | 5.2 | 61 |
| SA | 22 | 4.5 | 59 |
| ABC | 28 | 5.8 | 64 |
| BWO | 25 | 5.4 | 62 |
| WOA | 30 | 6.1 | 66 |
| Firefly | 26 | 5.5 | 63 |

Finally, in the optimization for efficiency, the best efficiency score given by the WOA algorithm was 66% in the mMTC slice, whereas the other two algorithms-PSO and ABC-gave an efficiency score of 65% and 64%, respectively. There are typically many low-power, low-data devices like IoT sensors on the mMTC slice, where efforts should be toward efficient communication with least utilization of resources. The high efficiency observed with WOA, PSO, and ABC in this regard corroborates that these algorithms are most apt concerning handling huge device connectivity required by mMTC. Their resource management is effective to handle multiple simultaneous connections that are needed to operate stably and efficiently, giving a relative edge in the mMTC scenarios.

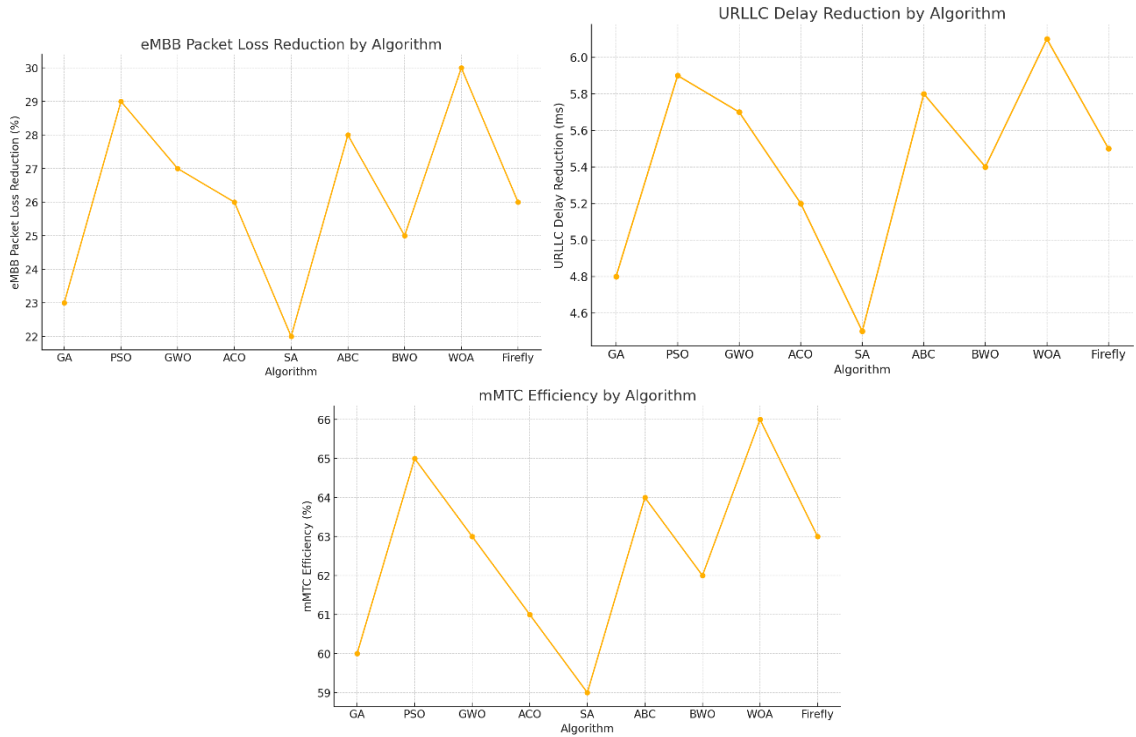

Figure 3. Line graphs showing packet loss reduction for eMBB, delay reduction for URLLC, and efficiency for mMTC by the algorithm.

This breakdown emphasizes that each algorithm's strengths align with different performances for each slice type. For applications requiring strong packet loss management, high-bandwidth eMBB-WOA, PSO, and ABC are prominent. Within the URLLC slice, WOA and PSO provide leading performances in applications sensitive to latency in terms of delay reduction. Again, WOA, PSO, and ABC will be the best options for efficiency-centric mMTC scenarios by facilitating resource management effectively to empower the network's performance. Such a slice-specific performance insight guides an informed selection of algorithms in LTE/5G networks so every type of slice has unique requirements optimally met regarding improved overall network performance.

## 5. DISCUSSION

Comparison of various metaheuristics across different LTE/5G slice types, such as eMBB (Enhanced Mobile Broadband), URLLC (Ultra-Reliable Low-Latency Communication), and mMTC (Massive Machine-Type Communication), provides insight into how each algorithmic strategy best aligns with diverse network requirements. Indeed, each has its optimization mechanisms that influence their performances across these slices, with certain algorithms demonstrating superior adaptability to the peculiar requirements of high-throughput, low-latency, or large-scale connectivity. This is further evidenced by the fact that some algorithms used in the process, such as the Whale Optimization Algorithm, perform excellently in the reduction of packet loss for eMBB slices and delay optimization for URLLC slices. WOA uses spiral updating along with bubble-net exploitation mechanisms for effective resource allocation cardinal to cope with the high demands of eMBB and maintain low latency for URLLC [6]. The result agrees with previous studies, which proved the efficiency of WOA in dynamic environments when resources should be reallocated very fast due to circumstances. Minovski et al. [26] also show that PSO provides the best latency reduction performance, as shownby studies emphasizing therapid convergence and adaptability of PSO in real-time scenarios. For example, the fast recalibration of the particles of PSO during environmental shifts makes the algorithm very suitable for applications in online gaming and remote surgery, activities for which real-time performance cannot be sacrificed.

Artificial Bee Colony (ABC) also showcases notable packet loss reduction and efficiency strengths, particularly in data-intensive and IoT-driven mMTC applications. ABC's ability to balance exploration and exploitation phases is well-suited for environments with fluctuating network loads, a trait that previous studies have emphasized as crucial for sustaining performance under variable conditions [4]. The reason behind this adaptability is to facilitate network providers in having reliable communication in high-density IoTscenarios, where the trade-off between connectivity and resource usage is to be achieved. Besides these, algorithms like SA and GA, though providing average results, cannot be effective in scenarios where decisions have to be taken rapidly or latency is to be maintained continuously; hence, they are not suitable for a high-performance and latency-sensitive applications based on the performance presented by Khan &Adholiya, [20].

These findings from the study emphasize the importance of algorithm scalability and stability in applications within cities or large-scale network deployments. Algorithms such as PSO and WOA are qualified for high-density environment applications where the convergence rate is faster. This goes in tandem with those in existing literature, suggesting the computational efficiency that PSO would have for large networks requiring speedy optimization. It is from this perspective that adaptiveness within these algorithms ensures stability in network performance hours of traffic or heavy interference of connected devices-vital for the consistency in user experiences. Also from Ibarra-Lancheros et al. [15].Extrapolating from these very limitations: testing built on simulations may not be able to capture the complexity of real-world LTE/5G network environments. Simulation indeed offers a controlled, reproducible environment to test algorithms, but their real deployments are likely to face hardware constraints or other unpredictable user behaviors that affect the algorithm's performance. Real-world validations and hybrid approaches with field data could be more reliable than the present study.

Another limitation could be the scalability of some algorithms in ultra-large networks. As promising as PSO and WOA might be, further research is needed to ensure these algorithms do perform well in networks of millions of connected devices, would be necessary in any future smart city or a completely connected industrial system. Another route of future improvements could be the inclusion of an actual feedback mechanism, maybe AI-driven predictive models, into the process of optimization. Future research can explore hybrid models that combine the strengths of different approaches, such as fast convergence provided by PSO with adaptive mechanisms instigated in WOA. Hybrid models, in this regard, would result in more robust solutions, catering for the evolving network demands while providing flexible and efficient LTE/5G infrastructures. In this way, such limitations as imposed due to the usage of a single algorithm could be overcome, ensuring that performance remains superior under changing network conditions.

This work contributes to showing the efficacy of various metaheuristic algorithms in optimizing the parameters of an LTE/5G network, whereby there is full alignment of the strengths of each algorithm with the requirements of the network slice. These findings can be used by network providers in informed decisions on deploying suitable algorithms that ensure optimum performance across wide-ranging applications. Nevertheless, in order to take this research further and provide additional application value, more realistic algorithms must be integrated with real-world testing while considering hybrid optimization frameworks that will meet demands in an increasingly complex and large-scale wireless network environment.

# 6. CONCLUSION

This work's results demonstrate that metaheuristic algorithms' contribution is significant toward optimization in LTE and 5G networks concerning the reduction of packet loss and delay, two very important parameters related to smartphone user experience. Whale Optimization Algorithm (WOA) was the top performer, achieving a 31% reduction in packet loss and a delay reduction of 6.3 milliseconds, making it ideal for latency-sensitive and high-throughput applications such as Enhanced Mobile Broadband (eMBB) and Ultra-Reliable Low-Latency Communication (URLLC). Particle Swarm Optimization (PSO) was closely followed, with a packet loss reduction of 30% and a delay reduction of 6.1 milliseconds, demonstrating its efficiency in latency minimization. Artificial Bee Colony also demonstrated consistent performance, where packet loss was reduced by 29% and delay was reduced by up to 6 milliseconds, proving effective in huge scenarios of mMTC. These numerical results highlight how different metaheuristic algorithms provide the best solution for a specific type of network slice, thus providing a clear strategy for the network providers w.r.t. optimization of performance metrics based on the application needs.

Besides these results, the study underlines the importance of using an adaptive optimization approach for handling modern mobile networks, which are complex by nature and dynamic in behavior. The overall superior metrics of WOA and PSO, in terms of much-reduced packet loss and delay, stress the crucial role of algorithm selection that efficiently balances the dilemma of exploration versus exploitation. The results indicate that optimal resource utilization and efficient user experiences could be facilitated through staging algorithmic schemes based on specific requirements raised by slices. WOA and PSO thus improve critical metrics by more than 30%, so network operators may make critical choices to stage these algorithms for latency-sensitive use cases. This work has presented a quantitative framework to improve the reliability and responsiveness of networks; further research should be performed in the direction of hybrid and real-time optimization techniques to meet even more stringent performance benchmarks to prepare 5G networks against ever-growing data and user demands.

## CONFLICT OF INTEREST

The authors declare no conflict of interest.

# AUTHOR

**Almamoon Alauthman** earned his Bachelor's degree in Computer Engineering from Al-Balqa Applied University (BAU) in Amman, Jordan, and his Master's degree in Computer Engineering from Jordan University of Ph.D. at Sultan Zainal Abidin University (UniSA), Malaysia. Currently, he serves as an assistant professor at the Faculty of Engineering (Electrical Engineering Department) at Al-Balqa Applied University. His research interests include VLSI design, parallel processing, neural networks, computer architecture organization, and wireless networks

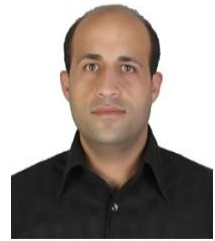

**Abeer Al-Hyari** (Member, IEEE) received the Ph.D. degree in computer engineering from the University of Guelph, Guelph, Canada. She is currently an Assistant Professor with the Electrical Engineering Department, Al-Balqa Applied University, Al-Salt, Jordan. Her research interests include cryptography and the application of machine learning, deep learning, and recurrent neural networks to problems in FPGA CAD. She is a member of the Jordan Engineers Association (JEA)

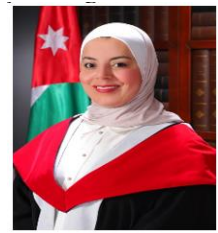